\documentclass[preprint, authoryear]{elsarticle}

\usepackage{natbib}
\usepackage[utf8]{inputenc} 
\usepackage[T1]{fontenc}
\usepackage{lmodern}
\usepackage[fleqn]{mathtools}
\usepackage{booktabs}
\usepackage{rotating}
\usepackage{amssymb}
\usepackage{amsmath}
\usepackage{graphicx}
\usepackage{caption}
\usepackage{epstopdf}
\usepackage{stackrel}
\usepackage{soul}
\usepackage{color}
\usepackage{float}
\usepackage[official]{eurosym}
\usepackage{tabularx}
\usepackage{multirow}
\usepackage{makecell}
\usepackage{lscape}
\usepackage{afterpage}
\usepackage{url}
\usepackage{mathrsfs}
\usepackage{chngcntr}
\usepackage{caption}
\usepackage{subcaption}
\usepackage{wasysym}
\usepackage{xfrac}
\usepackage[table]{xcolor}

\usepackage{ulem}
\usepackage{todonotes}

\usepackage{comment}

\newcommand{\Prob}{\text{Prob}}

\newdefinition{rmk}{Remark}
\newproof{pf}{Proof}
\newproof{pot}{Proof of Theorem \ref{thm2}}

\makeatletter
\def\ps@pprintTitle{%
	\let\@oddhead\@empty
	\let\@evenhead\@empty
	\def\@oddfoot{\centerline{\thepage}}%
	\let\@evenfoot\@oddfoot}
\makeatother

\begin{document}
\pagenumbering{gobble}
\title{Temperature Measurement in Agent Systems \\[12pt]
{\small(Version: \today)}}
%\tnoteref{t1,t2}}
%\tnotetext[t1]{This document is a collaborative effort.}
%\tnotetext[t2]{The second title footnote which is a longer.}

\author[hhu]{Christoph J.\ Börner}
\ead{Christoph.Boerner@hhu.de}	
\author[hhu]{Ingo Hoffmann\corref{cor1}}
\ead{Ingo.Hoffmann@hhu.de}	
%\author[hhu]{John H.\ Stiebel}
%\ead{John.Stiebel@hhu.de}

\cortext[cor1]{Corresponding author. Tel.: +49 211 81-15258; Fax.: +49 211 81-15316}
%\cortext[cor2]{Principal corresponding author}
%\fntext[fn1]{This is the specimen author footnote.}
%\fntext[fn2]{Another author footnote, but a little more longer.}
%\fntext[fn3]{Another author footnote, but a little more longer.}

\address[hhu]{Heinrich Heine University D\"usseldorf, Faculty of Business Administration \\ and Economics, Financial Services, 40225 D\"usseldorf, Germany,
\\ ROR: https://ror.org/024z2rq82}	

\begin{abstract}
Models for spin systems, known from statistical physics, are applied analogously in econometrics in the form of agent-based models.
The models discussed in the econophysics literature all use the state variable $T$, 
which, in physics, represents the temperature of a system.
However, there is little evidence on how temperature can be measured in econophysics, so that the models can be applied.
Only in idealized capital market applications has the relationship between temperature and volatility been demonstrated, allowing temperature to be determined through volatility measurements. The question remains how this can be achieved in agent systems beyond capital market applications.
This paper focuses precisely on this question. It examines an agent system with two decision options in a news environment, establishes the measurement equation, and outlines the basic concept of temperature measurement.
The procedure is illustrated using an example. 
In an application with competing subsystems, an interesting strategy for influencing the average opinion in the competing subsystem is presented.
\end{abstract}

\begin{keyword}
		 Agent System
	\sep Econophysics
	\sep Temperatur 							\\[6pt] 
	
	
	\noindent \textit{ORCID IDs:} 
	0000-0001-5722-3086 (Christoph J.~B\"orner), 
	0000-0001-7575-5537 (Ingo Hoffmann), 
	% 0000-0003-0088-2456 (John H.~Stiebel),    	\\[6pt] 
	
	%\noindent \textit{Acknowledgement:} The authors thank the editor and the anonymous reviewers for their helpful comments and constructive suggestions.
\end{keyword}
\maketitle
\newpage

%\pagenumbering{arabic} \setcounter{page}{1}
\pagenumbering{arabic}

% ---- Ab hier Textbody ----

\section{Introduction} \label{Introduction}
In the context of econophysics, methods from statistical physics are generally used to model the behavior of a large number of decision-makers (so-called agents) in a specific decision situation. For example, in capital market applications, the agents are investors who can buy or sell a specific stock and these methods help draw conclusions about the price movement of a stock and allows to develop a one-step ahead forecast model; see, e.g., \citet{Vikram.2011}. 

In the basic models it is assumed that the agents of the system have two possible decisions (Yes--No; Pro--Contra, Buy--Sell, Either--Or, True--False etc.) and are exposed to a news environment ${\cal B}$
\citep{Weidlich.1971, Kaizoji.2000, Michard.2005, Sornette.2006, Borghesi.2007,	Bouchaud.2013, Boerner.2023b, Boerner.2024}. 
The models then show that, given a fixed news environment ${\cal B}$ and a constant value of the state variable $T$, a surplus of decisions in the agent system develops in accordance with the news environment.
In a capital market example, a "buy on good news" behavior would predominantly be observed.
However, as the value of $T$ increases, this surplus decreases. This property is examined in various applications. Depending on the research field, the state variable $T$ is referred to by different names: temperature,
noise, irrationality, degree of randomness in agents’ decisions, the collective climate parameter or volatility; see, e.g., \citet{Weidlich.1971, Kaizoji.2000, Kozuki.2003, Oh.2007, Kleinert.2007, Kozaki.2008, Krause.2012, Bouchaud.2013, Crescimanna.2016, Boerner.2023b}. In the following, the state variable $T$ is referred to as temperature
and the question of measuring $T$ in an agent system is addressed. 
The answer to this question closes a research gap and elevates $T$ from a pure simulation parameter to a measurable state variable. 
The measurement of the state variable $T$ enables the transition from a qualitative to a quantitative evaluation of the econophysical models in the vicinity of the system's current state.

This paper follows the mainstream of research and also starts with the Ising model for spin systems \citep{Ising.1925}, known from statistical physics, as the basic model for the agent system, see, e.g., \citet{Weidlich.1971, Galam.1982, Chowdhury.1999, Kaizoji.2000, Eguiluz.2000, Bornholdt.2001, Krawiecki.2002, Sousa.2005, Sornette.2006, Zhou.2007, Zaklan.2008, Zaklan.2009, Sieczka.2011, Bouchaud.2013, Sornette.2014, Kaizoji.2015, Ko.2016, Bazart.2016, Crescimanna.2016, Smug.2018, Westphal.2020, Kopp.2022, Cividino.2023, Macy.2024} and the large amount of literature cited therein. 

The equation for measuring temperature is derived from the mean-field approximation of the Ising model \citep{Weiss.1907, Isihara.1971, Amit.1978, Landau.1980, Greiner.1995}, and it is shown that temperature measurement can be based on the surplus of decisions. Furthermore, it is shown that, similar to a thermometer, an accessible subsystem of the agent system can be used to infer the temperature of the entire agent system.
The concept of temperature measurement is described in detail and illustrated with an example. 
In a second application, 
the analysis of an idealized competing agent system illustrates how the decision surplus in such systems can be reduced by selectively altering a single parameter {\it c.p.}

This study is structured as follows. In the next section, temperature $T$ is defined as a state variable. In Section \ref{AgentSystem}, the model of a two-state agent system is introduced, and the central measurement equation for determining the temperature of the agent system is derived in Section \ref{TempMeasurement}. Applications are considered in Section \ref{Applications} and Section \ref{Limits} is dedicated to the limitations of the proposed method. The final section presents the conclusion.

\section{Temperature}\label{Temperature}
Analogical reasoning establishes a connection between energy $E$ in physics and utility $U$ in econophysics. While physical systems tend to minimize energy, econometric models assume utility maximization. Consequently, as a fundamental starting point for further calculations, the relationship $E = -U$ is frequently found in the econophysics literature
\citep{Marsili.1999, Sornette.2014, Boerner.2023c, Boerner.2023b}.

If the utility $U = U(S, {\bf X})$ is described as a function of the entropy $S$ and possibly other state variables $\bf X$ and the entropy is calculated from the microcanonical partition function $\Omega$ using the equation $S = k \ln\Omega$ \citep{Greiner.1995}, then \citet{Marsili.1999} suggests the definition of the state variable $T$ in econometrics:
\begin{flalign} \label{OekoTemp}
	T :=  - \left. \frac{\partial U}{\partial S}\right|_{{\bf X} = {\rm const.}}
\end{flalign}
Where $\bf X$ summarizes other (constant) variables, for example, the news environment $\cal B$ in the agent system. Equation (\ref{OekoTemp}) specifies that $T$ is not a simulation parameter but a state variable that describes the state of the agent system. This justifies the statement that an isolated agent system {\it has} a temperature. 
With this definition in Equation (\ref{OekoTemp}), it is possible in econophysics to build a consistent theory based on the methods of statistical physics.

Unlike in physics, $k$ ({\it phys.}: Boltzmann constant) is not a constant of nature in econophysics, but rather a predetermined parameter that measures the cost of information, see also \citet{Shannon.1948}.
From the perspective of an observer (such as a researcher, market analyst, investor, or reporter), $k$ quantifies the value of information about an agent's exact state in monetary units when the agent system is in a state of maximum uncertainty \citep{Boerner.2024}. 
For simplicity, we set $k = 1$ USD below.
Note: If utility $U$ is also measured in monetary units of the same currency, then it follows from Equation (\ref{OekoTemp}) that temperature $T$ is a unitless state variable in econophysics. This satisfies the dimensional equation, ensuring that the model framework is consistent with respect to units.

The relationship between energy and utility, along with Equation (\ref{OekoTemp}), provides a theoretical framework for interpreting the state variable $T$ in econophysics. Consequently, systems involving more complex phenomena, such as segregation 
\citep{Schelling.1971, Schelling.1978}, which challenge the analogy  $E = -U$
when considering additional individual utility 
\citep{Grauwin.2009, Lemoy.2011, Bouchaud.2013}, are not addressed here and are left for further research. 

\section{Two-State Agent System}\label{AgentSystem}
\subsection{Model}\label{Model}
This study focuses on the standard Ising model from physics for two spin states \citep{Ising.1925}, which is predominantly used in econophysics. Using the analogy $E = -U$, the utility for an observer of a specific configuration of the agent system is calculated as follows:
\begin{flalign}\label{utilityfunction}
		U  & =  +  \mu B  \sum\limits_{i} s_i 
		        +  J 		\sum\limits_{\langle ij \rangle} s_i s_j
\end{flalign}
if the system consists of $i = 1, \ldots, N$ agents.\\
\\
\noindent
The model variables and terms in detail:
\subsubsection*{Strengs of the news environment ${\cal B}$:}
In principle, the strength $B = |{\cal B}|$ and the type of a message (positive or negative) within a news environment can be measured through text analysis, see, e.g., \citet{Loughran.2016, Loughran.2020, Boerner.2023c}. The text analysis then provides an unitless value for ${\cal B}$ between $-1$ and $+1$. 
Therefore, the strength of the news environment satisfies $0\leq B \leq 1$. 
The following analysis will focus on cases in which the news environment establishes a clear preferential direction; accordingly, this paper concentrates on instances where $B > 0$.
\subsubsection*{Agent's behavior $s$:}
Agents can act in "conformity" to the news, $s = +1$, or in "non-conformity", $s = -1$.
In a capital market example of stock trading: If an investor (agent) buys on good news $({\cal B} > 0)$, this results in $s = +1$ ("conformity"). If the investor sells on good news, this results in $s = -1$ ("non-conformity"). If an investor sells on bad news $({\cal B} < 0)$, this results in $s = +1$ , and finally, if the investor buys on bad news, this results in $s = -1$.
\subsubsection*{Individual utility increase $\mu$:}
This model parameter is referred to differently in the literature depending on the application: willingness to adopt/buy, idiosyncratic judgment, see, e.g.,
\citet{Michard.2005, Sornette.2006, Borghesi.2007, Bouchaud.2013, Crescimanna.2016}.
In the general model, Equation (\ref{utilityfunction}), $\mu > 0$ represents the individual utility increase -- measured in monetary units -- triggered by an agent for an observer who behaves conformally ($s = +1$) to a news environment with maximum strength $B = |{\cal B}| = 1$. Unlike in physics, where parameter $\mu$ represents the magnetic moment, in econophysics, $\mu$ must be defined for specific applications or empirically measured, see, e.g., \citet{Boerner.2023c}.

In models of econophysics, the ratio $\sfrac{\mu}{k}$ frequently appears. Based on the previous explanations regarding the individual parameters, 
this ratio can, in a concrete application, be interpreted as a kind of fixed utility-cost ratio. Specifically, it represents the maximum achievable utility gain per agent conforming to the news environment, relative to the cost of obtaining precise information about an agent's decision state when the overall system is in a state of maximum entropy.
\subsubsection*{Term 1:}
This term of Equation (\ref{utilityfunction}) accounts for all gains and losses in utility resulting from the individual behavior $s$ of the agents within an external news environment ${\cal B}$ for a given configuration $(s_1, s_2, \ldots, s_N)$ of the agents.
\subsubsection*{Collective utility increase $J$:}
This model parameter measures the contribution of a pair of agents with the same
behavior to the utility, see, e.g., \citet{Bazart.2016}. In the general model, it is assumed that the utility for an observer increases when the agents behave similarly (both "conform" or both "non-conform"), such that $J$ is positive and measured in monetary units. The fundamental idea is that agents behave similarly when they perceive a utility gain in doing so. They adopt an observer role and adjust their behavior accordingly. In econophysics, this parameter must also be specified or empirically determined depending on the model's application, see, e.g., \citet{Boerner.2023}.
\subsubsection*{Term 2:}
In Equation (\ref{utilityfunction}) the notation $\langle ij \rangle$ denotes the summation over adjoining agents $j = 1, \ldots, z$ of an agent $i = 1, \ldots, N$ in the agent system, with $z\leq N$. Thus, the second term accounts for all gain and losses in utility resulting from the pairwise attitude of the agents for a given configuration $(s_1, s_2, \ldots, s_N)$ of the agents. Where each pair is counted only once. If two agents exhibit similar behavior, utility increases by $+J$; if they behave oppositely, utility decreases by $-J$.
\subsection{Mean-Field Approximation and Occupation Probabilities}\label{MeanField}
The utility function Equation (\ref{utilityfunction}) cannot generally be evaluated for arbitrary agent configurations concerning macroscopic variables, such as, for example, the decision surplus. However, special solutions for similar problems in one-dimensional \citep[$z = 2$]{Ising.1925} and two-dimensional \citep[$z = 4$]{Onsager.1944} lattice structures are well established in physics.
As in physics, econophysics typically addresses $d$-dimensional problems, where $z =2d > 4$, by approximating a mean field around an agent \citep{Weidlich.1971, Galam.1982, Brock.2001, Gordon.2005, Nadal.2005, Michard.2005, Gordon.2009, Bouchaud.2013}. This approach traces back to \citet[{\it phys.:} molecular field approximation]{Weiss.1907} and has become a standard method in both physics \citep{Isihara.1971, Amit.1978, Landau.1980, Greiner.1995} and econophysics, as demonstrated in the aforementioned studies. The core principle of the mean-field approximation is that an agent’s state is influenced by the average field generated by the states of its neighboring agents.

Following the original approach of \citet{Weiss.1907} the utility function Equation (\ref{utilityfunction}) can be written in mean-field approximation \citep{Greiner.1995}:
\begin{flalign}\label{MFutilityfunction}
			U & = + \mu B_{\rm eff}  \sum\limits_{i} s_i 
\end{flalign}
With the effective news environment:
\begin{flalign}\label{MFeffektivefield1}
	B_{\rm eff} & = B + \frac{1}{2}\frac{J}{\mu} z \; M
\end{flalign}
Where $M = \frac{1}{N} \sum_{i} s_i$ is the average surplus of decisions. Equation (\ref{MFeffektivefield1}) can be interpreted such that, for an individual agent, the strength $B$ of the news environment is amplified by the $z$ adjoining agents.

By summing over the so-called Boltzmann factors, which include the temperature $T$, the canonical partition function can be computed within the mean-field approximation, allowing for the determination of the Boltzmann-Gibbs distribution for the occupation probabilities \citep{Greiner.1995, Kaizoji.2000}. With $x = \frac{\mu B_{\rm eff}}{k T}$ the probabilities of each state  $s = (-1, +1)$ of an agent are:
\begin{flalign}\label{Probabilities} \nonumber
	P_{-} = \Prob(s = -1) 
	& = \frac{1}{1 +  \exp(+2x)} = \frac{N_{-}}{N}\\ 
	\\[-12pt] \nonumber
	P_{+} = \Prob(s = +1) 
	& = \frac{1}{1  + \exp(-2x)} = \frac{N_{+}}{N}
\end{flalign}
with $P_{-} + P_{+} = 1$ and $(N_{-}  \quad N_{+}) = (P_{-}  \quad P_{+}) \times N $ are the occupation numbers. A similar representation of the occupation
probabilities can be found in, e.g., \citet{Bouchaud.2013, Boerner.2023b}.

\section{Temperature Measurement}\label{TempMeasurement}
\subsection{Measurement Equation} \label{MeasurementEquation}
With the occupation numbers the average surplus of decisions is $M = \frac{N_{+}}{N} - \frac{N_{-}}{N} = P_{+} - P_{-}$ and $P_{\pm} = \frac{1}{2}(1 \pm M)$, see, e.g., \citet[Equation (44)]{Bouchaud.2013}.
Then, from Equation (\ref{Probabilities}) it follows:
\begin{flalign} \label{xDetermination} 
	2x & = \ln\left(\frac{P_{+}}{P_{-}}\right)  = \ln\left(\frac{1+M}{1-M}\right)
\end{flalign}
With $x = \frac{\mu B_{\rm eff}}{k T}$ and $B_{\rm eff}$ of Equation (\ref{MFeffektivefield1}) this leads to the measurement equation for temperature $T$:
\begin{flalign} \label{MeasurementEq}
	T &= \frac{2 \frac{\mu B}{k} + \frac{J z}{k}M}{\ln\left(\frac{1+M}{1-M}\right)}
\end{flalign}
With the message environment remaining constant in intensity, i.e.\ $B = $ const., the temperature $T$ becomes solely a function of the average surplus of decisions $M$. Once $M$ is determined, the current temperature $T$ of the agent system can be calculated using Equation (\ref{MeasurementEq}).

The measured values for $M$ in a capital market example are typically on the order of $10^{-2}$, see, e.g., \citet{Boerner.2023c}. If the Taylor expansion for $\ln\left(\frac{1+M}{1-M}\right) = 2M + \frac{2}{3} M^3 + \ldots $ is applied in Equation (\ref{MeasurementEq}), then for $M \approx 0^+$, the first-order approximation yields:
\begin{flalign}\label{TaylorMeasurementEq}
	T & = \frac{\mu B}{k} \; \frac{1}{M} + \frac{1}{2} \frac{J z}{k}
\end{flalign}
For a specific application, it is assumed that the model parameters $k, \mu, J$ 
and the average number $z$ of interacting agents are known.
Equation (\ref{MeasurementEq}) or (\ref{TaylorMeasurementEq}) then shows that a consistent temperature measurement in a constant news environment with strength $B$ can be obtained by determining the surplus of decisions $M$.

If there is hardly any benefit from the pairwise equilibrium behavior of agents, then in the limiting case $J \rightarrow 0$ the agent system is called an ideal agent system \citep{Bouchaud.2013, Boerner.2023b} and it follows for the temperature measurement:
\begin{flalign} \label{TempMeasuringIAS}
	T = \frac{\mu B}{k} \; \frac{1}{M}
\end{flalign}
Note: By rearranging the last equation for $M$ and differentiating with respect to $B$, Curie's law, well-known from physics, follows, taking into account that in physics the surplus of decisions $M$ is known as magnetization and is provided with an additional factor $\mu$ \citep[Equation (26.15)]{Fliebach.2018}.

\subsection{Characteristic Curves}\label{CharacteristicCurves}
\begin{figure}[H] %sonstige moegliche Einstellungen: [htbp]
	\centering
	\captionsetup{labelfont = bf, labelsep = none}
	\includegraphics[width=0.925\textwidth]{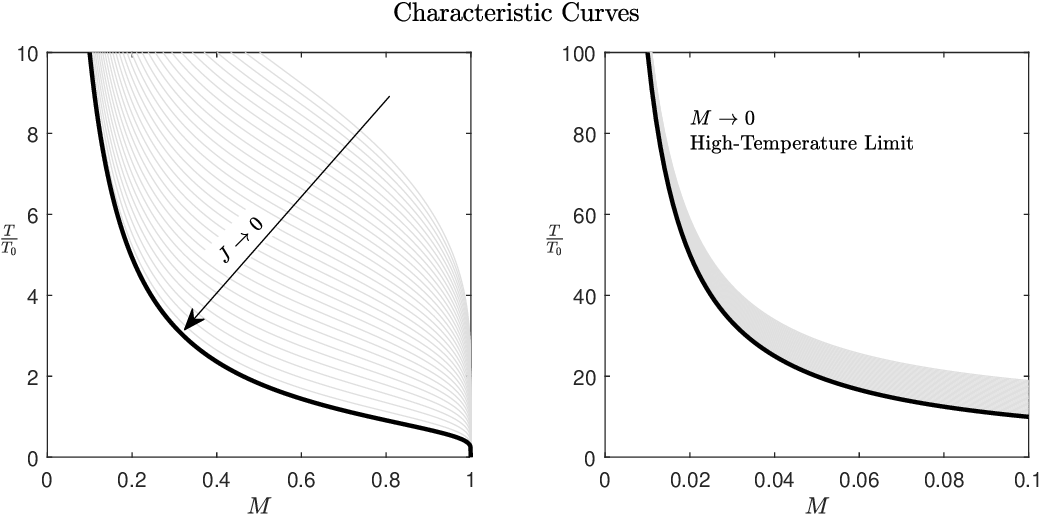} %Einstellung für ArXiv.org
	%\begin{quote}
	\caption[Mean Attitude]{\label{FigureCharacteristicCurves} 
	Set of characteristic curves for temperature measurement in an agent system. The right graph shows the characteristic curves for small $M$, so that the curves can be approximated by Equation (\ref{TaylorMeasurementEq}).}
	%\end{quote}
\end{figure}

Since $M$ is a unitless evaluation of the average surplus of decisions, the factor $\frac{\mu B}{k}$ defines a temperature scale for constant strength $B$ of the news environment and is abbreviated with $T_0 = \frac{\mu B}{k}$. Figure \ref{FigureCharacteristicCurves} shows the set of characteristic curves of the temperature measurement $\frac{T}{T_0}$, Equation (\ref{MeasurementEq}), as a function of $M$ for different utility increases, $J = 0\ldots 3$, from pairwise uniform behavior. The lower limit curve is reached at $J = 0$ for the ideal agent system. As an example, the figure shows the characteristic curves for $k = 1$ USD, $\mu = 1$ USD and $B =1$ with $z = 12$.
In capital market applications, values of $M$ are found in the order of $10^{-2}$ \citep{Boerner.2023c}.
It is conceivable that, in practical applications, most observations may fall within this range; thus, the graph on the right could be considered appropriate for use, where the characteristic curves exhibit a steep slope toward $M \rightarrow 0$, i.e.\ the high-temperature limit $T \rightarrow \infty$. 
In this region, distinguishing the characteristic curves becomes increasingly difficult. It can be observed that the influence of $J$ decreases, and many agent systems can likely be well approximated by an ideal agent system. Thus, the temperature of the agent system in many practical applications can be roughly determined using Equation (\ref{TempMeasuringIAS}).

\subsection{Measurement Procedure} \label{MeasurementProcedure}
In specific applications of agent-based models in capital markets, empirically observed volatility can be used to infer the temperature of the agent system, see, e.g., \citet{Boerner.2023b}. However, in arbitrary agent systems, volatility may not be readily measurable. This raises the question of how temperature can be determined as a state variable in these systems. Equation (\ref{MeasurementEq}) can now be used to define an appropriate measurement procedure in the corresponding category of models.

To determine the temperature $T$ of an arbitrary agent system, Equation (\ref{MeasurementEq}) must be evaluated, which requires the prior determination of $M$. 
This leads to the next question of how such a procedure might be formulated for these systems. The approach outlined below offers one possible solution.

The average surplus of decisions $M$ can be determined by a complete count of the $N$ individual decisions. The conforming, $N_{+}$, and non-conforming decisions, $N_{-}$, are then counted and offset against each other: $N_{+} - N_{-}$. A normalization with the factor $\sfrac{1}{N}$ then takes place and $M$ is thus determined. 
In general, however, a full assessment of all individual decisions is associated with considerable effort or may even be practically infeasible, which necessitates focusing on a suitable, representative sample when determining $M$, as is done in opinion polls, for example. With the appropriate subsample, the $M$ can only be determined approximately and the approximate temperature of the arbitrary agent system can be inferred using the measurement equation (\ref{MeasurementEq}). In a figurative sense, the subsample acts as a thermometer for measuring the state variable $T$. 

\subsection{Measurement Procedure -- Ideal Agent System}\label{MeasurementProcedureIAS}
In \citet{Boerner.2023b}, it is shown for the idealized agent system that the $N$ agents can be arranged linearly for illustrative purposes. This allows the sequence of their individual decisions $(s_1, s_2, \ldots, s_N)$ to be represented as a Bernoulli process. 
Bernoulli processes are strongly mixing and, therefore, ergodic \citep{Walters.1982}. The ergodicity implies that the spatial average (i.e., the mean across all agents) is equal to the time average (i.e., the mean of successive decisions of a single agent over time).

As a result, it is sufficient -- ceteris paribus -- to observe the decision behavior of a single representative agent over a sufficiently long period and record the temporal changes in their attitudes. Based on this time series, the quantities $N_{+}$ and $N_{-}$
(the number of conforming and non-conforming decisions, respectively) can be determined, from which the surplus $M$ follows.

An ideal agent system within the present modeling framework can thus be interpreted as a system of $N$ identical copies of a representative agent, in the sense of \citet[p.\ 109, 'representative particular']{Edgeworth.1881}
Accordingly, the temperature $T$ of the system can be assessed using Equation (\ref{MeasurementEq}), based on the observed behavior, $M$, of just one agent.

\section{Applications}\label{Applications}
The application described in this section shows how the temperature $T$ can be measured on an agent system. Furthermore, it is shown how a strategic action can be derived from a temperature equilibrium.

\subsection{Analysing Experimental Data from Human Decision Making Behaviour} \label{NeuroData}
The previously described procedure for determining the temperature is applied to the experiment carried out by \citet{Sun.2022}. In this experiment, the behavior of human participants in decision-making is investigated with the aim of empirically determining the parameters for a so-called drift-diffusion model, among other things. Such models are used in neuroscience, for example, to explain choice and response time data in perceptual decision making tasks. 

Of interest here for this paper is the experiment itself and the empirical results, which are described in Chapter III of Sun 2022's paper and in the caption of Figure 1 {\it op.\!\! cit.}\\

\noindent
Brief description of the experiment conducted by \citet{Sun.2022}:\\

Participants took part in a perceptual decision-making experiment where, in each round, they had to make a forced choice between two options (a so-called two-alternative forced-choice task).

\subsubsection*{The Task}
\begin{itemize}
	\item In each trial, participants were shown a Gabor patch (a visual pattern consisting of wavy lines – similar to a blurred stripe).
	
	\item This pattern was overlaid with dynamic visual noise to make the task more challenging.
	
	\item The participants’ task was to decide whether the pattern had a higher or lower spatial frequency (i.e., whether the lines appeared more closely spaced or more widely spaced).	
\end{itemize}
\subsubsection*{Difficulty}
\begin{itemize}
	\item The difficulty level was manipulated by changing the difference in spatial frequency between the two choices – sometimes the difference was large (easier), sometimes small (harder).
	
	\item There were three levels of difficulty, and participants completed multiple trials at each level.
\end{itemize}
\subsubsection*{Procedure}
\begin{itemize}
	\item Each participant completed 360 trials, organized into blocks with varying levels of difficulty.
	
	\item During the experiment, both behavioral responses (e.g., decision, reaction time) and brain activity were recorded using EEG with 128 electrodes.
\end{itemize}
\subsubsection*{Data}
\begin{itemize}
	\item EEG data were cleaned and then bandpass filtered to prepare them for the analysis conducted in \citet{Sun.2022}.
	
	\item The outcome of the decision as correct or incorrect.
\end{itemize}

The latter data are depicted in Figure 1 of Sun 2022's paper and serve as a basis for further processing in our analysis. 
The description of the experiment provides indications that participants are individually asked to make a choice (decision), so there is no coordinated behavior of the participants. 
This describes the prerequisite for an ideal agent system and $J = 0$ must be set.

In the following, it is assumed that the outcome of the individual decision -- correct (conform) or incorrect (non-conform) -- of each participant (agent) has the same value for the experimenters (observer) and set this to $\mu = 1$ USD. Similarly, it is assumed that for the experimenters the information about an agent's exact state when the agent system is in a state of maximum uncertainty has a fixed value and this is set to $k = 1$ USD.

In this experiment, the Gabor patch shown to the participants corresponds to the news environment $\cal B$. The noise imposed by the experimenters reduces the strength of the news environment. The data suggest that the data shown in Figure 1 of Sun 2022's paper for Choice 1 tend to originate from a stronger news environment (less imposed noise, thus proportionally fewer incorrect answers) and the data for Choice 2 originate from a slightly weaker news environment. For further analysis, we set the strength of the news environment $B_{\rm C1} = 1.0$ for Choice 1 and $B_{\rm C2} = 0.9$ for Choice 2. With these assumptions, $T_{0, {\rm C1}} = 1.0$ and $T_{0, {\rm C2}} = 0.9$ immediately follow.

To estimate the value of $M$, the respective areas under the curves in Figure 1 of Sun 2022's paper are evaluated, yielding an approximation of the corresponding $M$ values.
We counted the pixels with the freely available graphics tool GIMP and determined the values noted in Table \ref{TemperatureResults}.

\begin{table}[H]
	\footnotesize \noindent \centering
	\begin{tabularx}{\textwidth}{l|X|XX|X|X} 
		\toprule
		Choice
		& \multicolumn{1}{l|}{Scale}
		& \multicolumn{2}{l|}{Pixel}
		& \multicolumn{1}{l|}{$\o$ Surplus}
		& \multicolumn{1}{l}{Temperature}	   			\\ [2pt]
		No.\
		& $T_0	$
		& $N_{+}$
		& $N_{-}$
		& $M    $				
		& $T    $											\\[2pt]

		\midrule\midrule  %% Bis hier KOPFZEILE der Tabelle
		1 
		& 1.0
		& 5869
		& 2805
		& 0.353
		& 2.709									     \\[2pt]
		{}
		&
		& (blue full)
		& (red doted)
		&
		& 											\\
		\midrule %% Trennlinie zwischen den einzelnen Nachrichten  
		2 
		& 0.9
		& 6042
		& 3137
		& 0.316
		& 2.746							     \\	[2pt]
		{}
		&
		& (red full)
		& (blue doted)
		&
		& 					\\     
		\bottomrule
	\end{tabularx}
	\caption{: Analysis of the experiment conducted by \citet{Sun.2022} with regard to the measurement of temperature $T$. The termperature is calculated with Equation (\ref{MeasurementEq}). For the parameters, see the main text. The colors refer to Figure 1 of Sun 2022's paper.}
	\label{TemperatureResults} 
\end{table}
Temperature constitutes a state variable and is therefore an intrinsic property of the agent system itself. 
It is, in the model category considered here, independent of the Gabor patches presented to the participants (agents) during the experiment. Thus, one would expect the temperature to be the same for both experiments.
The above analysis indicates that approximately the same temperature was measured for both experiments. Therefore, the current temperature of $T = 2.73$ with a scattering width of $\pm0.03$ would be used for this ideal agent system in further model applications. The latter are beyond the scope of this study and left for future research.

\subsection{Influencing Others}\label{Influencer}
An idealized theoretical application is considered to explore a potential strategy for weakening a competing agent system with regard to the average surplus of decisions that conform to the news environment.

The starting point is an arbitrary agent system consisting of many agents, which is conceptually divided into two parts by a partition wall.
The both resulting agent systems share the same news environment, ${\cal B} = {\cal B}_1 = {\cal B}_2$, identical per-agent benefits for conformity, $\mu = \mu_1 = \mu_2$, equal costs for acquiring information about the state of a single agent under maximum uncertainty, $k = k_1 = k_2$, and the same assumed number of neighbors, $z =z_1 = z_2$.

The only heterogeneity characteristic is that $J_1 = 0$ and $J_2 = J $. This means that the first agent system is an ideal agent system, and in the second agent system, the agents are guided not only by the news environment but also by the attitudes of the other agents.
It is further assumed, that the average surplus of the decisions in the two agent systems is small, $0 < M_1, M_2 \ll 1$, so that the Equations (\ref{TaylorMeasurementEq}) and (\ref{TempMeasuringIAS}) can be applied. 

Assuming an equilibrium $T = T_1 = T_2$ ({\it phys.}: thermal equilibrium), then:
\begin{flalign}\label{ConnectionEq} \nonumber
	\frac{\mu B}{k} \; \frac{1}{M_1}  
	& = \frac{\mu B}{k} \; \frac{1}{M_2} + \frac{1}{2} \frac{J z}{k} \\
	M_2 
	& = M_1 \frac{1}{1-\frac{1}{2} \frac{J}{\mu}  \frac{z}{B} M_1}
\end{flalign}
If the Taylor expansion $\sfrac{1}{(1-x)} = 1 + x + x^2 + \ldots$ for $ x \approx 0$
is used on the right-hand side of the last line of equation (\ref{ConnectionEq}), the first-order approximation is:
\begin{flalign}\label{ConnectionEq2}
		M_2 & = M_1 + \frac{1}{2} \frac{J}{\mu}  \frac{z}{B} M_1^2
\end{flalign}
Equation (\ref{ConnectionEq2}) shows that in equilibrium of the agent systems, the average surplus of decisions is smaller in the ideal agent system than in the non-ideal agent system: $M_2 > M_1$. However, this equation also shows that the average surplus $M_2$ can be reduced in the second agent system as soon as it becomes possible to increase the utility $\mu$ per agent that behaves in conformity to the news environment. 
The agents of the first agent system would therefore have to strive to increase the universally valid parameter $\mu$ in order to weaken the second agent system with respect to the state variable $M_2$.
For large $\mu$, the importance of micro-alliances among agents disappears, and both systems exhibit approximately the same (smaller) average surplus of decisions.
This relationship could have been guessed, but here it follows from theory.
\section{Limits} \label{Limits}
Possible limitations of applicability have already been discussed in various sections. In particular, the restrictions resulting from the analogy (physics to econophysics) were addressed in Section \ref{Temperature}. There are further potential shortcomings of the approach that will be briefly summarized here. For a more detailed discussion of the limitations, see, e.g., \citet{Boerner.2023}. All described limitations may serve as a promising basis for further research.

\subsection*{Utility Function and Model}
Numerous extensions of the utility function defined in Equation (\ref{utilityfunction}) are conceivable and highlight the limitations of the current approach. In line with mainstream literature, we assume homogeneous agent parameters, setting $\mu$ and $J$ equal for all agents. 
While more complex variants incorporating distributions of these parameters -- such as $\rho(\mu)$, individual $\mu_i$, or asymmetric $J_{ij}$ values -- have been proposed in sociophysics and related domains, see, e.g., \citet{Foley.1999, Durlauf.1996, Castellano.2009}, they are not considered here.

Similarly, we model the agent system using square or hypercube lattice structures with binary configurations, consistent with much of the literature, see, e.g., \citet{Bornholdt.2001, Sieczka.2008}. Configurations with more possible states for the agents, higher-order interactions or more complex spatial configurations (e.g., agent systems with triple states, hexagonal lattices or $U\sim s_{i_1} s_{i_2} s_{i_3} \ldots s_{i_n}$) are also beyond the scope of this study.

\subsection*{Mean-Field Approximation and Number of Adjoining Agents}
The mean-field approximation is justified in many contexts and has been extensively tested and empirically validated in the field of physics. 
As in physics, however, the reliability of its applicability in econophysics is highly sensitive to the number $z$ of neighboring agents considered. In this study, it is assumed that the number of interacting agents is sufficiently large to justify the approximation.

Neither the influence of the number of neighboring agents nor the potential impact of fluctuations in the effective news environment ${B}_{\rm eff}$ -- particularly when $z$ is small and neighboring agents exhibit significant variability in their decision-making -- has been examined here. These fluctuations become negligible in the limit $z \rightarrow \infty$ and can thus be ignored \citep{Amit.1978}. However, for very small values of $z$, discrepancies between theoretical predictions and empirical observations are likely to emerge.

\subsection*{Taylor Expansion}
The derivations and conclusions are essentially based on the assumption that the average surplus of decisions, $M$, is small, making a first-order approximation via Taylor expansion justifiable. A small $M$ can presumably be assumed in many practical applications. However, as $M$ increases, at least a second-order approximation should be considered, which may lead to additional or potentially modified conclusions. This aspect was not further investigated in the present analysis.

\subsection*{Random Variable $M$}
In general, $M$ is a random variable. \citet{Boerner.2023b} analyzes this in the context of an ideal agent system and demonstrates that $M$ follows a normal distribution in this case, with variance inversely proportional to the number of agents, i.e., $\propto \sfrac{1}{N}$, and thus approaching zero as $N \rightarrow\infty$. When the number of agents is very large, fluctuations in $M$ become negligible; see \citet{Fliebach.2018}, and compare also \citet{Greiner.1995}.

In physics, this assumption is justified, as systems often consist of on the order of $10^{24}$ particles. In econophysics, however, the number of agents is usually many orders of magnitude smaller, which means that fluctuations in $M$ can become significant and may affect temperature measurements. The latter is especially true if only samples are considered to determine $M$.

For further research, it would be interesting to determine what the distribution function looks like in general and how fluctuations affect temperature measurements when the number of agents involved is finite.

\subsection*{Non-Equilibrium Considerations}
The models considered here apply to a single point in time and, analogous to physics, assume equilibrium. Temporal dynamics such as memory effects, changes in individual preferences, or developments in habitus are not captured \citep{Bouchaud.2013, Boerner.2023c}. Addressing these aspects would require non-equilibrium thermodynamic approaches, see, e.g., \citet{Isihara.1971}.

\subsection*{Negative Temperature}
In rare cases, the agent system may exhibit predominantly non-conforming behavior with respect to the news environment, i.e., $M<0$. Such an inversion is known in statistical physics and is associated with the assignment of a negative temperature to the inverted state. These states are generally unstable and tend to relax within a short period of time \citep[p.\ 223]{Greiner.1995}. The same assumption is made for the agent system considered here: an inverse state -- possibly triggered by two consecutive news items with rapidly changing signs -- is expected to relax quickly.

The transition from the inverted to the normal state typically proceeds through a series of non-equilibrium states, which require modeling approaches from non-equilibrium thermodynamics; see, e.g., the treatment of propagating partition functions in \citet{Isihara.1971}.

The preceding sections focused on the "normal" case, in which the agent system predominantly aligns with the news environment. As such, system inversions and their relaxation dynamics were not considered here.

\section{Conclusion} \label{Conclusion}
A specific class of agent system models was examined, grounded in the foundational spin system model introduced by \citet{Ising.1925}, which is well-known in physics. A key state variable emerging from these models is temperature. However, existing literature offers little guidance on how this variable can be measured in arbitrary agent systems.
This study addresses that gap by proposing a measurement equation that makes the state variable temperature fundamentally accessible. The approach is based on assessing the balance of conforming and non-conforming decision tendencies. This evaluation can be performed for the entire agent system or based on a representative sample.

The applicability of the measurement method was illustrated through a neuroscience experiment. 
An idealized application demonstrated 
how the uniqueness of collective decision-making in a competing agent group can be strategically weakened.

By identifying the current temperature as the system’s instantaneous operating point, this framework allows for analyzing how small temperature changes affect system behavior in the vicinity of that point, e.g.\ in a sensitivity analysis.

All models have limitations. The assumptions and boundaries of the models presented here have been thoroughly examined and critically discussed. These limitations pave the way for future research aimed at refining and extending the framework.

\bibliography{../../020_Literatur/005_CitaviBibTexFile/CitaviHerding}
\end{document}